\documentclass[a4paper,12pt]{article}
\usepackage{amsmath,graphicx,subfigure}       
\renewcommand{\L}{\Lambda}
\def\a{\alpha}
\def\b{\beta}
\def\c{\chi}
\def\g{\gamma}
\def\d{\delta}
\def\e{\epsilon}
\def\m{\mu}
\def\n{\nu}
\def\r{\rho}
\def\p{\psi}
\def\s{\sigma}
\def\no{\nonumber}
\def\oh{\frac{1}{2}}
\def\pa{\partial}

\begin{document}
\begin{titlepage}
\begin{flushright}
IFUM 656/FT \\
March 2000 \\
\end{flushright}
\vspace{1.5cm}
\begin{center}
{\bf \large AXIAL CURRENT IN QED AND SEMI-NAIVE DIMENSIONAL RENORMALIZATION }
{\bf \large }
\footnote{Work supported in part by M.U.R.S.T.}\\
\vspace{1 cm}
{ M. PERNICI} \\ 
\vspace{2mm}
{\em INFN, Sezione di Milano, Via Celoria 16, I-20133 Milano, Italy}\\
\vspace{0.6 cm}
{ M. RACITI}\\
\vspace{2mm}
{\em Dipartimento di Fisica dell'Universit\`a di Milano, I-20133 Milano,
Italy}\\
{\em INFN, Sezione di Milano, Via Celoria 16, I-20133 Milano, Italy}\\
\vspace{2cm}
\bf{ABSTRACT}
\end{center}
\begin{quote}
We renormalize at two loops the axial current and $F \tilde{F}$ in
massless QED, using the recently proposed semi-naive dimensional
renormalization scheme. 
We show that the results are in agreement with those in the
Breitenlohner-Maison-'t Hooft-Veltman scheme, previously obtained 
indirectly by making a three-loop computation.

\end{quote}
\end{titlepage}
\section*{Introduction}
Dimensional regularization \cite{HV,DR} and minimal subtraction
\cite{tH} are the most convenient tools for multi-loop computations.
However the accuracy of the computations of chiral quantities is much
lower than for non-chiral quantities, due to the difficulties
encountered in restoring the chiral symmetries in the BMHV scheme  
\cite{HV,BM}, the only one known until recently
to deal consistently with $\g^5$ in dimensional regularization. 

Few chiral quantities have been computed in the BMHV scheme; among the
most accurate computations made up to now there are 
the renormalization at two loops of the non-singlet axial current \cite{VL}, 
of the singlet axial current and of $F \tilde{F}$ \cite{Larin} in
QCD, and the corresponding calculation of the three-loop anomalous
dimension of the singlet axial current \cite{Larin}.

The techniques used in these papers to guarantee
the validity of the chiral Ward identities in the BMHV scheme are
indirect, and not suitable to be generalized to chiral gauge theories.
The most impressive trick used is the determination of two-loop finite
counterterms in the fermionic sector of the singlet axial current operator by
making a three-loop computation in its gluonic sector \cite{Larin}.

Recently it has been shown \cite{P} that there is a consistent
 extension of the BMHV
scheme, called semi-naive dimensional regularization (SNDR),
 in which after minimal subtraction 
only few graphs can produce terms breaking the
chiral Ward identities. The SNDR scheme has been applied in that paper to
 the case of the renormalization at two loops of the Yukawa model in
 presence of external gauge fields, previously studied in the BMHV
 scheme \cite{PRR}.

As a preliminary investigation in the use of the techniques developed
in \cite{PRR} and \cite{P} for renormalizing gauge theories with
chiral interactions,
in this letter we apply the SNDR scheme to renormalize at two loops the
axial current and $F \tilde{F}$ in QED; to check
the chiral Ward identities and to compute the anomalous dimensions we
will use the Wilsonian methods introduced in \cite{PRR}.

We find that using the minimal subtraction prescription in SNDR (MS-SNDR)
\cite{P} for the axial current the anomalous axial Ward identity 
is preserved by choosing a non-minimal subtraction for $F \tilde{F}$.
Vice-versa, one can satisfy the anomalous axial Ward identity by
choosing the minimal subtraction for the operator $F \tilde{F}$ and 
a non-minimal subtraction for the axial current.
With the latter renormalization prescription we find the same three-loop
anomalous dimension of the axial current as in the BMHV scheme
\cite{Larin}, where minimal subtraction is made on $F \tilde{F}$.
Notice that in the BMHV it is not possible to choose minimal
subtraction on the axial current, due to the presence of the same kind
of spurious anomalies appearing in the non-singlet axial current.

In the first section we review the Adler-Bardeen theorem \cite{AB},
following to a large extent the regularization-independent derivation
in \cite{BMS}.

In the second section we review the SNDR scheme and we perform the
two-loop renormalization of the axial current and of $F \tilde{F}$ in
QED.

In the third section we discuss our methods of computation, we
compare our results with those in
\cite{Larin} and we discuss in this context the relation 
between SNDR and BMHV.

\section{Review of the Adler-Bardeen theorem}

 The classical action of QED with $N_f$ massless fermions in presence
 of sources for $J_{\m}^5$ and $K_\m$ is
\begin{eqnarray}\label{cl}
S^{(0)} =\int \bar{\psi} \g_\m \pa_\m \psi + \frac{1}{4} F_{\m\n}^2 +
\frac{1}{2 \alpha} (\pa_\m V_\m)^2 + i e V_\m \bar{\psi} \g_\m \psi + 
A_\m J_{\m}^5 + \chi_\m K_\m
\end{eqnarray}
where we define
\begin{eqnarray}\label{ops}
J_{\m}^5  \equiv i \bar{\psi} \g_\m \g^5 \psi ~~;~~
K_\m \equiv 4 i \e_{\m\n\r\s} V_\n \pa_\r V_\s
\end{eqnarray}
We use Euclidean space conventions.

On the functional generator $\Gamma = \Gamma[V,\p,\bar{\p},A,\c]$
of $1PI$ vertex functions the
renormalization group equation reads
\begin{eqnarray}\label{rge}
&&{\cal{D}} \Gamma = 0  \\
&&{\cal{D}} \equiv \m \frac{\pa}{\pa \m} + \b \frac{\pa}{\pa e} +
\d \a \frac{\pa}{\pa \a} - \sum_i \g_i N_i -
\int (\g_{\c A} \c_\m \frac{\d}{\d A_\m} +
\g_{A \c} A_\m \frac{\d}{\d \c_\m}) \no
\end{eqnarray}
where the index $i$ runs over 
$V_\m, \psi, \bar{\psi}, A_\m, \c$; $\g_i$ and $N_i$ are
the corresponding anomalous dimensions and number operators;
$\g_{A\c}$ and $\g_{\c A}$ are the mixing 
anomalous dimensions for the operators (\ref{ops}).

The vectorial Ward identity reads
\begin{eqnarray}\label{vector}
&&G_v \Gamma = - \frac{1}{\a} \pa^2 \pa_\m V_\m +
4 i \e_{\m\n\r\s} \pa_\m \c_\n \pa_\r V_\s 
 \no \\
&&G_v \equiv \pa_\m  \frac{\d}{\d V_\m} + i e \p  \frac{\d}{\d \p} -
i e \bar{\p}  \frac{\d}{\d \bar{\p}}
\end{eqnarray}
Due to the linearity of the breaking terms, they do not need to be
renormalized. The proof of this fact in the case of the gauge-fixing
term \cite{Zinn} can be straightforwardly extended to the $\c$ term.

The axial Ward identity is
\begin{eqnarray}\label{axial}
&&G_a \Gamma = \rho \pa_\m \frac{\d \Gamma}{\d \c_\m} \no \\
&&G_a \equiv \pa_\m  \frac{\d}{\d A_\m} + i \g^5 \p  \frac{\d}{\d \p} +
i \bar{\p} \g^5  \frac{\d}{\d \bar{\p}}
\end{eqnarray}
The system of constraints (\ref{rge},\ref{vector},\ref{axial})
satisfies the following consistency conditions:
\begin{eqnarray}
\label{consve}&&[{\cal{D}} , G_v] = 
i \b (\p  \frac{\d}{\d \p} - \bar{\p}  \frac{\d}{\d \bar{\p}}) +
 \g_V \pa_\m  \frac{\d}{\d V_\m} \\
\label{consax}&&[{\cal{D}} , G_a - \rho \pa_\m \frac{\d }{\d \c_\m}] =
(\g_A- \r \g_{\c A}) \pa_\m \frac{\d }{\d  A_\m} +
(\g_{A\c} - \r \g_\c - \b \frac{\pa \r}{\pa e}) 
\pa_\m \frac{\d}{\d \c_\m}  \no
\end{eqnarray}
which, together with eq.(\ref{vector}) imply respectively
\begin{eqnarray}
\label{ve}&&\b = e \g_V ~~;~~\d = \g_\c = -2 \g_V ~~;~~ \g_{A \c} = 0 \\
\label{ax}&&\g_A = \r \g_{\c A} ~~;~~ \g_\c = - \b \frac{\pa ln \r}{\pa e}
\end{eqnarray}
from which it follows that, since $\beta \neq 0$,
\begin{eqnarray}
\r = c a
\end{eqnarray}
where  $a \equiv \frac{e^2}{16 \pi^2}$ and
$c$ is a constant in $e$ which is fixed by the one-loop anomaly
computation to be 
\begin{eqnarray}
c = -N_f 
\end{eqnarray}
This proof of the Adler-Bardeen theorem \cite{AB}, which is close 
to the one in \cite{BMS}, relies on the validity of (\ref{vector});
this relation can be imposed independently of the regularization scheme.
In a regularization which
respects the vectorial Ward identities, like Pauli-Villars or the
BMHV and the SNDR  dimensional regularization schemes, 
this relation implies that the term $\chi_\m K_\m$ of the bare
action does not need to be renormalized.

The first relation in (\ref{ax}) gives $\g_A$
at $l$ loops in terms of $\g_{\c A}$ at $(l-1)$-loops.

Allowing for finite renormalization $f_\a$ and $f_\c$ respectively
for the gauge-fixing and $\chi_\m K_\m$ terms, the relations
(\ref{ve}) are modified in a trivial way
\begin{eqnarray}
\d = - 2 \g_V + \b \frac{\pa ln f_\a}{\pa e} ~~;~~
\g_\c = - 2 \g_V + \b \frac{\pa ln f_\c}{\pa e}
\end{eqnarray}
Using (\ref{ax}) one gets 
\begin{eqnarray}
\r  = \frac{c a}{f_\c}
\end{eqnarray}

\section{Axial current in SNDR}
Let us review the BMHV and the SNDR dimensional regularization schemes. 

In the BMHV scheme \cite{BM} one considers the Lorentz covariants  
$\delta_{\mu \nu}$, $\g_\m$, $p_\m$, etc. as formal objects, satisfying the
usual tensorial rules.   
 $\delta_{\mu \nu}$ is the Kronecker delta in
$d=4-\epsilon$ dimensions; a formal rule for summed indices is
given:

\begin{equation}
\delta_{\mu \nu} \delta_{\nu \r} = \delta_{\m\r} ~~;~~
\delta_{\mu \nu} p_{\nu} = p_{\mu}~~;~~\delta_{\mu \mu} = d 
\end{equation}

The gamma `matrices' $\gamma_{\mu}$  satisfy the
relation
\begin{equation}
\{ \gamma_{\mu},\gamma_{\nu} \} = -2 \delta_{\mu \nu} I
\end{equation}
where $I$ is the identity.
The trace is cyclic and satisfies
\begin{equation}\label{trace}
tr~ I = 4 
\end{equation}

In the BMHV scheme additional `$(d-4)$-dimensional' or `evanescent'
tensors $\hat{\delta}_{\mu \nu},\hat{p}_{\mu}$ and $\hat{\gamma}_{\mu}$ 
are introduced; the Kronecker delta in the $(d-4)$-dimensional space
is $\hat{\delta}_{\mu \nu}$ , satisfying 
\begin{eqnarray}
&&\hat{\delta}_{\mu \nu}  \delta_{\nu \rho} = 
\hat{\delta}_{\mu \nu}  \hat{\delta}_{\nu \rho} = 
\hat{\delta}_{\mu \rho}~~;~~
\hat{\delta}_{\mu \mu} = - \epsilon \nonumber \\
&&\hat{p}_{\mu} \equiv \hat{\delta}_{\mu \nu} p_{\nu} ~~;~~
\hat{\gamma}_{\mu} \equiv \hat{\delta}_{\mu \nu} \gamma_{\nu} 
\end{eqnarray}
The Kronecker delta in four dimensions in $\bar{\delta}_{\m\n}$,
satisfying
\begin{eqnarray}
\bar{\delta}_{\m\n} \equiv \delta_{\m\n} - \hat{\delta}_{\mu \nu}
 ~~;~~
\bar{p}_\m  \equiv \bar{\delta}_{\m\n} p_\n  ~~;~~
\bar{\g}_\m  \equiv \bar{\delta}_{\m\n} \g_\n
\end{eqnarray}

The Levi-Civita antisymmetric tensor has no evanescent component:
\begin{equation}
\hat{\delta}_{\mu \nu} \epsilon_{\nu \rho \sigma \tau} = 0 
\end{equation}

$\g^5$ is defined by
\begin{equation}\label{g5}
\g^5 = \frac{1}{  4!} \epsilon_{\mu \nu \rho \sigma} \gamma_{\mu}
\gamma_{\nu} \gamma_{\rho} \gamma_{\sigma}
\end{equation}
which implies
\begin{eqnarray}\label{g51}
\{ \g^5, \gamma_{\mu} \} = 2 \hat{\gamma}_{\mu} \g^5 ~~;~~ (\g^5)^2 = I
\end{eqnarray}
The fact that $\g^5$ is not-anticommuting with $\g_\m$ leads to
violations of the (non-anomalous) axial Ward identities. On the other
hand the possibility of having anomalous axial currents must be
contemplated, so that it is necessary to define $\g^5$ so that it is
not anticommuting. However (\ref{g51}) introduces many breaking terms
which are not related to anomalies; these terms are sometimes called
spurious anomalies \cite{Trueman}.

The SNDR scheme \cite{P} is an extension of this scheme. 
Add to the BMHV Dirac algebra the objects $\eta$ and $\eta_1$
 satisfying the following defining relations:
\begin{eqnarray}\label{sndr}
&&\{ \eta, \gamma_{\mu} \} = 
\{ \eta_1, \gamma_{\mu} \} = 2 \hat{\gamma}_{\mu} \eta_1   \no    \\
&&\eta^2 = I ~~;~~ 
\eta \eta_1 = \eta_1 \eta = \eta_1^2 ~~;~~ \eta_1^3 = \eta_1 \no \\ 
&&tr~ \eta \gamma_{\mu} \g_\n \g_\r \g_\s = 
tr~ \eta_1 \gamma_{\mu} \g_\n \g_\r \g_\s = 
4 ~\epsilon_{\mu \nu \rho \sigma} \\
&&tr~ \eta_1 =  tr~ \eta_1^2 = 0~~;~~
tr~ \eta_1^2 \g_{\m_1} ... \g_{\m_r} = 0 \no \\
&&tr~\eta_1 \g^5 \g_{\m_1} ... \g_{\m_r} = 
tr~\eta \g^5 \g_{\m_1} ... \g_{\m_r} = 
tr~\g_{\m_1} ... \g_{\m_r} \no
\end{eqnarray}
The trace is cyclic on this enlarged algebra;
$\eta_1^2$ is a projector.

The idea of this regularization is that using $\eta$ instead of $\g^5$
in the tree-level chiral vertices, the number of spurious anomalies is
greatly reduced. Let us discuss separately the cases of open and
closed fermionic lines belonging to a $1PI$ Feynman graph.

i) {\bf Open fermionic lines }

In an open fermionic line $\eta$
can be anticommuted naively modulo terms belonging to the $\eta_1^2$
subspace, i.e. monomials containing $\eta_1$; algebraic manipulations
on a monomial containing $\eta_1$ give again terms in the $\eta_1^2$
subspace, which cannot be confused with those belonging
to the orthogonal subspace.  
The minimal subtraction on a $1PI$ Feynman graph
 consists in subtracting all the  
 poles in $\e$ and all the finite terms containing $\eta_1$.
In \cite{P} it is explained why the subtraction of these finite terms
is necessary and why it can be considered to be minimal.
The idea is that in the process of removing the regulator $\eta$ is
homomorphically mapped in $\g^5$;
since $tr~(\g^5 - \eta)\eta_1 = 4$ , there is no way to extend this
trace-preserving homomorphism to $\eta_1$. Therefore the $\eta_1$
terms must be subtracted before applying this homomorphism.
Notice that the necessity of subtracting the terms in the $\eta_1^2$
subspace has nothing to do a priori with the requirement of satisfying
the chiral Ward identities, but it has to do simply with the inner
consistency of SNDR.

In the BMHV scheme $\g^5$ can be anticommuted naively modulo terms
containing again $\g^5$; the latter terms can generate spurious anomalies,
which cannot be distinguished from the
non-anomalous contributions, so that their subtraction is non-trivial,
unlike in SNDR.

ii) {\bf Closed fermionic lines }

If in a trace there is no $\eta_1$ and an even number of $\eta$,
these $\eta$ can be anticommuted naively.
 If in a fermionic trace there is at least one $\eta_1$, all the
$\eta$'s   in that trace can be replaced with $\eta_1$; if at this point
there is an even number of $\eta_1$'s, the trace vanishes; these
properties reduce greatly the possibility of occurrence of spurious anomalies.

 If in a fermionic trace there is
an odd number of $\eta$ and $\eta_1$, they can be replaced with $\g^5$.  
The true anomalies originate from these traces.

\vskip 1 cm

The bare action in SNDR corresponding to the classical action
(\ref{cl}) is 
\begin{eqnarray}\label{bare}
&&S = \int Z_\p 
\bar{\psi} \g_\m [\pa_\m + i e V_\m + i A_\m \eta ] \psi + 
\frac{1}{4} Z_V F_{\m\n}^2 + \frac{1}{2 \alpha} (\pa_\m V_\m)^2 + \no \\
&&i A_\m \bar{\psi} \g_\m (Z_{A2}\eta_1 +Z_{A3} \g^5) \psi + 
\chi_\m (K_\m + Z_{\c A} i \bar{\p} \g_\m \g^5 \p )
\end{eqnarray}
where the sources satisfy $\hat{A}_\m = 0 = \hat{\c}_\m$.
The bare action in MS-SNDR is vector gauge invariant apart from tree-level
terms; in this scheme the vectorial Ward identities are automatically
satisfied.
\begin{figure}
\centering
\subfigure[]{\label{fig:1:a}\includegraphics[width=3cm]{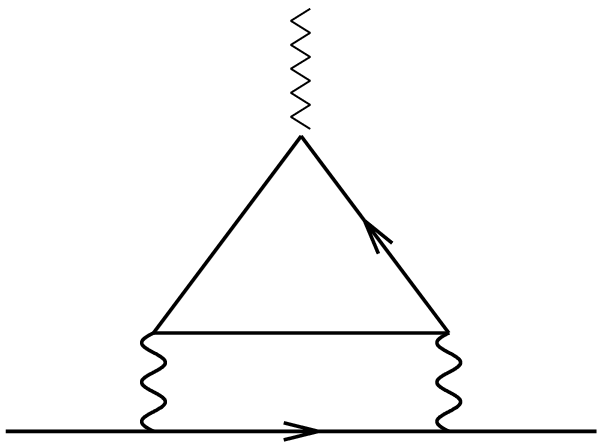}}
\hspace{0.4cm}
\subfigure[]{\label{fig:1:b}\includegraphics[width=3cm]{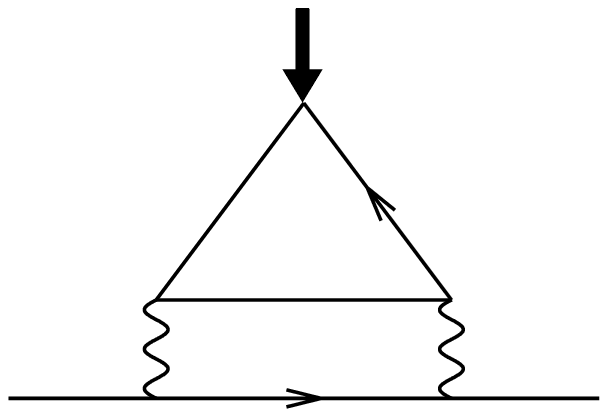}}
\hspace{0.4cm}
\subfigure[]{\label{fig:1:c}\includegraphics[width=3cm]{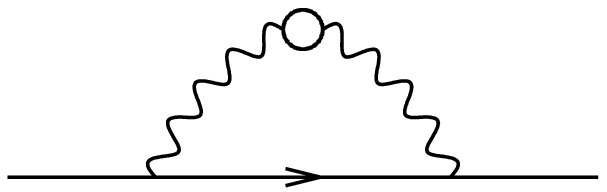}}
\caption{(a) $\Gamma_{\bar{\p}\p A \m}$ ; 
(b) corresponding ${\cal{T}}_a$ contribution;
(c) $\Gamma_{\bar{\p}\p \c \m}$.
These are the only graphs for which the axial Ward identity
is not trivially satisfied at two loops in SNDR.}
\label{fig:1}
\end{figure} 
 We have computed the two-loop
renormalization constants such that the axial Ward identity
(\ref{axial}) is satisfied.
The only axial Ward identity which is not trivially satisfied 
is the one involving Figure \ref{fig:1},
leading to the non-minimal renormalization 
\begin{eqnarray}\label{nonmin}
Z^{non-MS}_{A3} - \r Z^{non-MS}_{\c A} = 3 N_f a^2 
\end{eqnarray}
All the other renormalization constants can be chosen minimal in SNDR.
At the first two loops the renormalization constants are, for $\a = 1$,
\begin{eqnarray}\label{z}
&&Z_V = Z_V^{MS} = 1 - \frac{8}{3 \e} N_f a - \frac{4}{\e} N_f a^2 \no
\\
&&Z_\p = Z_\p^{MS} = 1 - \frac{2 a}{\e} +(\frac{2}{\e^2} + \frac{3}{2 \e} + 
\frac{2 N_f}{\e}) a^2 \no \\
&&Z_{A2} = Z_{A2}^{MS} =- 4 a + (-\frac{16}{3 \e} + \frac{4}{9}) N_f a^2 + 
(\frac{8}{\e} + 22)a^2  \\
&&Z_{A3} = \frac{12}{\e} N_f a^2  +Z^{non-MS}_{A3} \no \\
&&Z_{\c A} = - \frac{24 a}{\e} + 
(-\frac{64}{\e^2} + \frac{16}{3 \e}) N_f a^2 +
(\frac{48}{\e^2} + \frac{84}{\e}) a^2 + Z^{non-MS}_{\c A} \no
\end{eqnarray}
where $Z^{non-MS}_{A3}$ is of order $a^2$, whereas 
$Z^{non-MS}_{\c A}$ is considered at order $a$ only, since its $a^2$
term is related by the axial Ward identity to the three-loop
$a^3$ term of $Z^{non-MS}_{A3}$.
 
The corresponding anomalous dimensions are given by
\begin{eqnarray}
&&\g_V = \frac{4}{3} N_f a + 4 N_f a^2 \no \\
&&\g_\p = a - (2 N_f + \frac{3}{2}) a^2  \\
&&\g_{\c A} = 24 a - 72 a^2 -\frac{32}{3} N_f a^2 +
\frac{16}{3} N_f a Z^{non-MS}_{\c A} \no
\end{eqnarray}
the remaining anomalous dimensions being fixed by eqs.(\ref{ve},
\ref{ax}).
In particular one determines in this way indirectly the three-loop
anomalous dimension of the axial current
\begin{eqnarray}
\g_A = -24 N_f a^2 + 72 N_f a^3 + \frac{32}{3} N_f^2 a^3 -
\frac{16}{3} N_f^2 a^2 Z^{non-MS}_{\c A}
\end{eqnarray}

\section{Discussion of the results}
To obtain the results in the previous section we have used the
Wilsonian method devised in \cite{PRR} for computing in a systematic
way the finite counter\-terms needed to restore chiral Ward identities
in dimensional regularization schemes. 

In this approach the renormalization of the theory is obtained
imposing renormalization conditions on a Wilsonian functional 
$\Gamma^\Lambda$, which is perturbatively defined with the same
Feynman rules for the vertices as in the usual $1PI$ functional
generator $\Gamma$, but with the usual propagators $D$ replaced by 
`hard' propagators 
\begin{eqnarray}
D^H = (1-K^\Lambda) D ~~;~~
K^\Lambda(p) = (\frac{\Lambda^2}{p^2 + \Lambda^2})^2
\end{eqnarray}

Let us denote by $S_W$ the local functional whose tree-level part is
equal to the classical action (\ref{cl}), and which for $l \geq 1$ is
equal to the marginal part of the Wilsonian functional $\Gamma^\L$;
one has
\begin{eqnarray}\label{sw}
&&S_W = \int a_\p \bar{\psi} \g_\m \pa_\m \psi +
\oh a_{V1} (\pa_\m V_\n)^2 + \oh a_{V2} (\pa_\m V_\m)^2 + 
\frac{1}{4} a_{V4} V_\m^2 V_\n^2 + \no \\
&&i a_{\p V} V_\m \bar{\psi} \g_\m \psi +
A_\m (a_{A\c} K_\m  + a_{A} J_{\m}^5) + \c_\m (a_\c K_\m + a_{\c A} J_{\m}^5)
\end{eqnarray}
The renormalization group equation and the Ward identities assume now
the `effective' form
\begin{eqnarray}
\label{wilsrg}&& {\cal{D}} S_W = {\cal{T_\g}} \\
\label{wilsve}&&\int \e_v G_v S_W = \int \e_v (- \frac{1}{\a} \pa^2 \pa_\m V_\m +
4 i \e_{\m\n\r\s} \pa_\m \c_\n \pa_\r V_\s ) + {\cal{T}}_v  \\
\label{wilsax}&&\int \e_a G_a S_W = \int \e_a \rho \pa_\m \frac{\d
S_W}{\d \c_\m} + 
{\cal{T}}_a 
\end{eqnarray}
where the ${\cal{T}}$ terms are contributions to the number operators
and to the contact term operators at the Wilsonian scale.
The Green functions of the ${\cal{T}}$ terms can be computed using 
Feynman rules as described in \cite{PRR}.

We use the hard propagator $D_{\m\n}^H = D^H \d_{\m\n}$ for the photon
in the Feynman gauge. 

To compute the renormalization group functions we use (\ref{wilsrg}).
The derivative with respect to the gauge-fixing
parameter in (\ref{wilsrg}) in $\alpha =1$ is treated
as the insertion of the operator $\int \oh (\pa_\m V_\m)^2$.

As discussed before, the vectorial Ward identities are trivially
satisfied in MS-SNDR, so that we have to discuss only the axial Ward
identities (\ref{wilsax}).

One has 
\begin{eqnarray}
{\cal{T}}_a = \int \pa_\m \e_a
 (b_\p J_{\m}^5 + b_\c K_\m )
\end{eqnarray}
The axial Ward identity gives the following relations
\begin{eqnarray}
&&\label{wax1} a_A - a_\p + b_\p - \r a_{\c A} = 0 \\
&&\label{wax2} a_{A \c} + b_\c - \r a_{\c} = 0
\end{eqnarray}
The marginal part of $S_W$ exists for $\L > 0$; we will compute its
coefficients at $\L = \m$, the dimensional regularization scale.

At zero and one loops one has, in the MS-SNDR scheme,
\begin{eqnarray}\label{oneloop}  
&&a_\p^{MS} = 1 + \frac{a}{6} ~~;~~
a_A^{MS} = 1 - \frac{43 a}{60}   \no \\
&&a_{\c A}^{MS} =  \frac{77 a}{5} ~~;~~
a_{A\c}^{MS} = - \frac{2 N_f a}{3}  ~~;~~a_{\c} = 1 \\
&&b_\p = \frac{53 a}{60} ~~;~~ 
b_\c = - \frac{N_f a}{3}  \no
\end{eqnarray}
The axial Ward identities (\ref{wax1},\ref{wax2})
are manifestly satisfied in the SNDR scheme as long as $\g^5$ or the
Levi-Civita tensor does not appear in the Feynman rules.
This is the case at one loop, with the exception of the 
$\Gamma_{\bar{\p}\p \c \m}$
vertex, which is related by the axial Ward identity 
to the two-loop insertions of $\Gamma_{\bar{\p}\p A \m}$, so that
we will discuss it later.

To illustrate this renormalization procedure, consider 
for instance the relevant part of the unrenormalized Wilsonian axial vertex 
$\Gamma_{\bar{\p}\p A \m}^{unren \L}$ at one loop:
\begin{eqnarray}  
i a [(\frac{2}{\e}- \frac{43}{60}) \bar{\g}_\m \eta + 
4 \bar{\g}_\m \eta_1 ]
\end{eqnarray}
The pole term and the $\eta_1$ terms are minimally subtracted, giving 
the renormalization constants $Z_\p$ and $Z_{A2}$ at one loop
(\ref{z}); 
the remaining finite part, evaluated in four dimensions 
(i.e. $\eta \to \g^5$), gives the contribution to $S_W$ in 
(\ref{sw},\ref{oneloop}).

At two loops the MS-SNDR scheme satisfies manifestly the axial Ward
identities, apart from the contribution of Fig.1.
The two-loop Wilsonian Green function 
$\Gamma_{\bar{\p}\p A \m}^{unren \L}$, 
unrenormalized at two loops but renormalized at one loop,
 gives the following relevant contributions
\begin{eqnarray}
&&i a^2 \bar{\g}_\m \{ 
-(\frac{2}{\e^2} + \frac{3}{2 \e} - \frac{806251}{204120} +
\frac{12937 v}{4374}) \eta -
(\frac{8}{\e} + 22) \eta_1 - \no \\
&&N_f (\frac{2}{\e} - \frac{1363}{5670}  + \frac{1136 v}{243}) \eta +
N_f (\frac{16}{3 \e} - \frac{4}{9} ) \eta_1  -
N_f (\frac{12}{\e} + \frac{2449}{270} + \frac{1072 v}{81}) \g^5 \} \no
\end{eqnarray} 
where the $\bar{\g}_\m \g^5$ contribution, which comes from the graph in
Figure \ref{fig:1:a}, is the only one for which minimal subtraction is not
automatically sufficient to preserve the axial Ward identity.
The pole terms and the $\eta_1$ terms are minimally subtracted, giving 
the renormalization constants $Z_\p$ ,$Z_{A2}$ 
and $Z_{A3}^{MS}$ at two loop (\ref{z}).

\begin{figure}
\centering
\subfigure[]{\label{fig:2:a}\includegraphics[width=2cm]{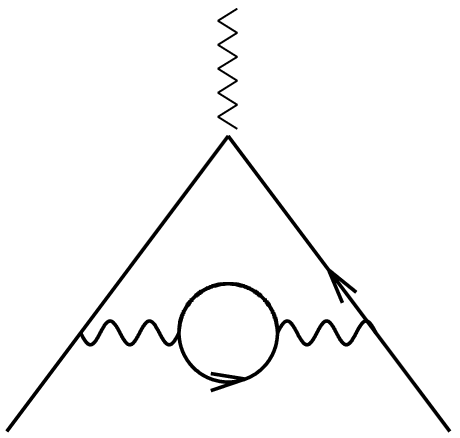}}
\hspace{0.4cm}
\subfigure[]{\label{fig:2:b}\includegraphics[width=2cm]{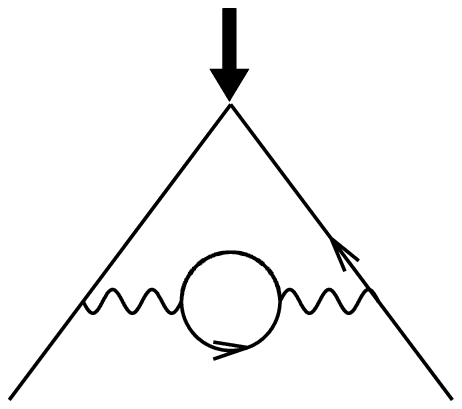}}
\hspace{0.4cm}
\subfigure[]{\label{fig:2:c}\includegraphics[width=3.5cm]{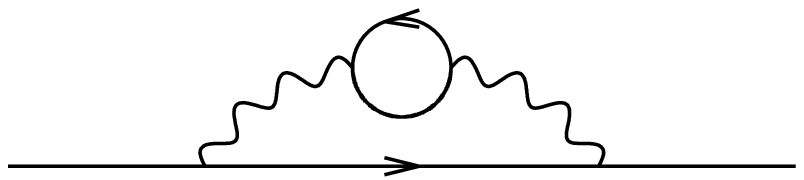}}
\caption{Example of graphs for which the axial Ward identity is
trivially satisfied at two loops in SNDR.}
\label{fig:2}
\end{figure}

Let us consider for instance the contributions to the axial Ward
identity due to the graphs in Figure \ref{fig:2}
and the corresponding counterterms:
\begin{eqnarray}  
&&a_A^{MS}(fig.2) = ( \frac{1363}{5670}-  \frac{1136 v}{243})  N_f a^2
 \no \\
&&b_\p^{MS}(fig.2) = - ( \frac{28393}{17010} + \frac{2272 v}{729})
N_f a^2 \no \\
&&a_\p^{MS}(fig.2) = -(\frac{1736}{1215} + \frac{5680 v}{729}) N_f a^2 
 \no
\end{eqnarray}
There is no $K_\m$ insertion corresponding to the graphs in Figure
\ref{fig:2}, so that $a_{\c A}^{MS}(fig.2) = 0$ and
the axial Ward identity (\ref{wax1}) is satisfied as expected.

Let us consider finally the only axial Ward identity which 
is not automatically satisfied
in this scheme. For the graph in Fig. 1 one has
\begin{eqnarray}  
&&a_A^{MS}(fig.1) = -(\frac{2449}{270} + \frac{1072 v}{81}) 
N_f a^2 \no \\
&&b_\p^{MS}(fig.1) = (-\frac{2519}{270} +\frac{1072 v}{81})
N_f a^2 \no \\
&&a_\p^{MS}(fig.1) = 0 \no 
\end{eqnarray}
The corresponding $K_\m$ insertion gives 
$a_{\c A}^{MS}(fig.1) = \frac{77a}{5}$ (see eq.(\ref{oneloop}))
so that in the minimal scheme the 
axial Ward identity (\ref{wax1}) is not satisfied
\begin{eqnarray}  
(a_A - a_\p + b_\p - \r a_{\c A})^{MS}(fig.1) = 
-3 N_f a^2
\end{eqnarray}
and the non-minimal counterterms in (\ref{nonmin})
must be added to restore (\ref{wax1}).

To obtain the bare action in the BMHV scheme, one can use the values
found for $S_W$ as renormalization conditions at the Wilsonian scale.
The result is simply phrased: it is sufficient to
replace $\eta$ and $\eta_1$ in (\ref{bare}) with $\g^5$.
In fact, if $\eta$ or $\eta_1$ belongs to an open fermionic line of a
$1PI$ graph, anticommuting it through the gamma matrices of the
fermionic lines one uses only the relations in the first line of (\ref{sndr}),
which agree with (\ref{g51}) after replacing $\eta$ and $\eta_1$ with
$\g^5$.
If $\eta$ or $\eta_1$ belongs to an open fermionic line of a $1PI$
graph, it gives the same as in the case in which  it is replaced
by $\g^5$, due to the last line in (\ref{sndr}).

The correspondence between the bare actions in these two schemes is not
always so simple; in presence of more than  one chiral vertex 
the BMHV scheme produces many
finite counterterms which are absent in the SNDR scheme; the
difference is due to the fact that $tr ~ \eta_1^2 = 0$ in SNDR (see
(\ref{sndr})) whereas $tr~(\g^5)^2 = 4$ in BMHV; 
see for instance the Yukawa model in the BMHV scheme \cite{PRR} and in
the SNDR scheme \cite{P}.

Our results agree with those
in \cite{Larin}, after the changes due to conventions,
provided one makes the minimal choice for $Z_{\c A}$.
Similarly for the anomalous dimensions.
In the BMHV scheme it is possible to make this minimal choice, whereas one
cannot make minimal subtraction of the axial current, since the
one-loop finite counterterm 
$\int 4 a A_\m J_{\m}^5$ in the bare action
necessary to satisfy the axial Ward identity
 cannot be replaced by the term $\int \frac{4}{N_f} \c_\m J_{\m}^5$
without violating the classical limit of the operator $K_\m$.

On the other hand in the SNDR scheme it is possible to make the minimal
subtraction either on $K_\m$ or on $J_{\m}^5$, to all orders in
perturbation theory.

Let us review the tricks used in \cite{Larin} to perform this
computation.
To compute the one-loop finite counterterm for the axial current, 
comparison between
the axial and the vector vertices is made, as
suggested in \cite{Trueman}. To compute the two-loop finite 
counterterm the same trick cannot be used in the case of the singlet
axial current, due to fact that
the graph in Figure \ref{fig:1} has no counterpart in the vector vertex.
To fix this finite counterterm, instead of checking directly the axial Ward
identity on the axial vertex at two loops, the three-loop computation
of $<\pa_\m J_{\m}^5 V_\n V_\r>$ has been made in \cite{Larin}, 
obtaining these two-loop
finite terms by consistency with the Adler-Bardeen theorem.

\section{Conclusion}
The SNDR scheme is a consistent extension of the BMHV 
dimensional regularization and renormalization scheme, which has been
introduced to reduce the number of spurious anomalies present in the
latter scheme. 

As a preliminary investigation in gauge theories with chiral couplings,
in this letter we have applied the SNDR scheme to the renormalization of
the axial current in QED.  In this case there is only one chiral
vertex, so that there are few spurious anomalies, which have been
determined in the BMHV scheme in \cite{Larin}, together with the
three-loop anomalous dimension of the singlet axial current in QCD.
We find agreement with these results in the QED case. 
The correspondence between the
SNDR and the BMHV scheme is in this case so easy that it can be made
even at the bare action level. As expected, we found that it is easier to
satisfy the axial Ward identity in the SNDR scheme than in the BMHV
scheme. These computational advantages are expected to be much greater
in the case of chiral gauge theories, 
which have been renormalized systematically in the  BMHV scheme only
at one loop \cite{Martin}.


\begin{thebibliography}{99}
\bibitem{HV} G. 't Hooft and M. Veltman, Nucl. Phys. {\bf B44} (1972) 189.
\bibitem{DR} C.G. Bollini and J.J. Giambiagi, Phys. Lett. {\bf 40B}
(1972) 566; \\
J.F. Ashmore, Nuovo Cimento Lett. {\bf 4} (1972) 289; \\
G.M. Cicuta and E. Montaldi, Lett. Nuovo Cimento {\bf 4} (1972) 329;
\\
D.A. Akyeampong and R. Delburgo, Nuovo Cimento {\bf 17A} (1973) 578.
\bibitem{tH} G. 't Hooft, Nucl. Phys. {\bf B61} (1973) 455.
\bibitem{BM} P. Breitenlohner and D. Maison, 
Comm. Math. Phys. {\bf 52} (1977) 11.
\bibitem{VL} S.A. Larin and J.A.M. Vermaseren, Phys. Lett. {\bf B259}
(1991) 345.
\bibitem{Larin} S.A. Larin, Phys. Lett. {\bf B303} (1993) 113. 
\bibitem{P}  M. Pernici, `Semi-naive dimensional renormalization',
hep-th/9912278.
\bibitem{PRR}  M. Pernici, M. Raciti and F. Riva,
`Dimensional renormalization of Yukawa theories via Wilsonian
methods', preprint  hep-th/9912248, to appear on Nuclear Physics B.
\bibitem{AB} S. Adler and W. Bardeen, Phys. Rev. {\bf 182} (1969)
1517.
\bibitem{BMS} P. Breitenlohner, D. Maison and K.S. Stelle,
Phys. Lett. {\bf 134B} (1984) 63.
\bibitem{Zinn} see e.g. J. Zinn-Justin, `Quantum Field theory and 
Critical Phenomena', Clarendon Press, Oxford (1989).
\bibitem{Trueman} T.L. Trueman, Phys. Lett. {\bf 88B} (1979) 331 ;
 Z. Phys. {\bf C 69} (1996) 525.
\bibitem{Martin} C.P. Martin and D. Sanchez-Ruiz, `Action principles,
restoration of BRS symmetry and the renormalization group equation for
chiral non-Abelian gauge theories in dimensional renormalization with
a non-anticommuting $\g^5$' , preprint hep-th/9905076. 
\end{thebibliography}
\end{document}